\documentclass{article}

\usepackage{arxiv}

\usepackage[utf8]{inputenc} 
\usepackage[T1]{fontenc}    
\usepackage{hyperref}       
\usepackage{url}            
\usepackage{booktabs}       
\usepackage{amsfonts}       
\usepackage{nicefrac}       
\usepackage{microtype}      
\usepackage[T1]{fontenc}
\usepackage{times}
\usepackage[pdftex]{graphicx} 
\usepackage{amsthm}
\usepackage{amsmath}
\usepackage{flexisym}
\usepackage[numbers]{natbib}

\newtheorem{definition}{Definition}

\title{Subjective Knowledge and Reasoning about Agents in Multi-Agent Systems}

\author{
  Shikha Singh \\
  Department of Computer Science and Engineering\\
  IIT Madras \\
   \And
 Deepak Khemani \\
 Department of Computer Science and Engineering\\
 IIT Madras \\
}

\date{}

\begin{document}
\maketitle

\begin{abstract}
Though a lot of work in multi agent systems is focused on reasoning about knowledge and beliefs of artificial agents, an explicit representation and reasoning about presence/absence of agents, especially in the scenarios where agents may be unaware of other agents joining in or going offline in a multi-agent system, leading to partial knowledge/asymmetric knowledge of the agents is mostly overlooked by the MAS community. Such scenarios lay the foundations of cases where an agent can influence other agents' mental states by (mis)informing them about presence/absence of collaborators or adversaries. In this paper we investigate how \textit{Kripke structure based} epistemic models can be extended to express the above notion based on an agent's subjective knowledge and we discuss the challenges that come along.
\end{abstract}


\section{Introduction} 
The research community has successfully used Kripke models to model epistemic situations in multiagent systems. But it seems 
necessary to have a broader notion of epistemic models that may not always be defined within the interpretation of the structural properties 
of S5 or KD45 Kripke models which are used to represent uncertainty (knowledge) of agents in a multi-agent system. We envisage an underlying 
global epistemic structure in a multi agent system, wherein the agents may have access to (non)overlapping local regions. Hence the agents 
can be said to possess \textit{subjective knowledge} about the system. Such a global epistemic structure allows us to talk about the possibility of agents being ignorant of the 
presence of other agents. While we talk of ignorance, we point out that there are two levels of it as we perceive it: one, where an agent is 
uncertain whether something is true or false but it is certain that it exists to be known, while the other is at the level of existence 
itself. We restrict ourself to modeling the first level of ignorance only. The community uses KD45 Kripke structures to model \textit{beliefs} instead of knowledge but these beliefs are typically \textit{about the world} and not about the 
\textit{agents}. The agents are assumed to be \textit{aware of} all other agents.\newline

In this paper we explore ways to relax this assumption in the epistemic models based on KD45 Kripke structures. First we motivate the work with an example:
\emph{Consider a scenario involving multi-vehicle search and rescue. Three (unmanned) vehicles V1, V2, V3 have been assigned to survey a sequence 
of points looking for survivors. Each vehicle knows whether there are survivors at their respective surveyed points and send updates to each 
other. This is how they are pre-scripted to collaborate. Meanwhile V1 fails (assuming a failure signal be sent out to others\footnote{we proceed with this assumption for the rest of the paper}), V2 and V3 
have to complete the task on their own. The belief bases of V2 and V3 should get updated so that the strategy to 
accomplish the remaining task can be recomputed. If on the other hand, more vehicles (which were off-line or disconnected earlier) join the team 
midway or say 
V1 is up and running after certain interval of time, the existing ones should be able to `understand and update' its presence in their belief 
bases to recompute 
their strategies}.\newline 
Consider another setting where the system modeler may want the agents to work independently, as if they are in a single 
agent setting, in the initial phases of the task and collaborate with each other at a later stage. Such scenarios can arise in privacy-
preserving use cases. We feel that a formalism that facilitates an artificial agent to explicitly model and reason with its (un)certainty about 
other agents will be useful in the settings discussed above. We explore other avenues as well where the capability of reasoning about agency 
can be used by an agent to influence other agents' beliefs about agency. We discuss this in detail in a separate section dedicated to ontological lies\footnote{the term `ontological lies' refers to lies about presence/absence of an agent}.

\section{Background}
Before discussing the scope of extending epistemic logic to support our formalism we briefly look at the basics. Epistemic logic, the logic  of knowledge and beliefs \cite{hintikka1962knowledge}, is used to reason with an agent's knowledge about the world as well as its own (and the others') beliefs about the world, beliefs about beliefs and so on.\newline
 \textbf{Language}: Let $\mathcal{P}$ and $\mathcal{AG}$ be a finite set of propositions and agents respectively. The language can be constructed using the formulae given below:
\[\phi = p\, |\, \neg \phi\, |\, (\phi \wedge \varphi)\, |\, K_i\phi\] 
where $p \in \mathcal{P}$ and $i \in \mathcal{AG}$. The intended interpretation of $K_{i}\phi$ is `agent i knows $\phi$'. It is the basic system of knowledge known as $\mathcal{S}5$ system whose axiomatization will be discussed after we have looked at its \textit{possible worlds} semantics using Kripke models.


\textbf{Semantics}: Kripke structures enable the agents to think in terms of \textit{possible worlds}  \textit{accessible} to them and an agent is said to know or believe some state of affairs to be true, if and only if, it is the true with respect to all the possible worlds accessible to it.
\begin{definition}[Kripke Model]
Given the set of propositions $\mathcal{P}$ and the set of agents $\mathcal{AG}$, a Kripke model is a triple $M = \langle S,R,V\rangle$, where: S is a set of states, $R$ is a function, such that for all $i \in \mathcal{AG}$, there is an accessibility relation defined $R(i) \subseteq S \times S$, $V: P \to 2^S$ is a valuation function for all $p \in \mathcal{P}$, the set $V(p) \subseteq S$ is the set of states in which p is true.
\end{definition}
$M = \langle S,R,V \rangle$ is called an epistemic model and all the relations $R$ in $M$ are equivalence relations (which explains the truth and introspective properties, i.e. $S$5 properties, of knowledge). Epistemic formulas are interpreted on pairs $(M,s)$, also called as a pointed model, and a formula $\phi$ is true in $(M,s)$ is written as $(M,s) \models \phi$. Thus, the satisfaction of a formula can be expressed as: (i)  $(M,s) \models p \text{ iff } s \in V(p)$, (ii) $(M,s) \models (\phi \wedge \varphi) \text{ iff } (M,s) \models \phi \text{ and } (M,s) \models \varphi$, (iii) $(M,s) \models \neg \phi \text{ iff  not } (M,s) \models \phi$ and (iv)  $(M,s) \models K_i \phi \text{ iff  for} \text{ all $t$ such that } (s,t) \in R(i), (M,t)\models \phi$. Here, $(s,t) \in R(i)$ stands for existence of accessibility relation of agent \textit{i} between the two epistemic states \textit{s} and \textit{t}.\newline

\textbf{Axiomatisation}: The axioms of $S$5 system of knowledge include all instantiations of propositional tautologies, along with the following axioms:
\begin{itemize}
\item Distribution of $K_i$ over $\to$: $K_i(\phi) \wedge K_i(\phi \to \varphi) \to K_i(\varphi)$
\item Modus Ponens: From $\phi$ and $\phi \to \varphi$, infer $\varphi$
\item Necessitation of $K_i$: From $\phi$, infer $K_i \phi$
\item Truth: $K_i \phi \to \phi$
\item Positive Introspection: $K_i(\phi) \to K_i(K_i(\phi))$
\item Negative Introspection: $\neg K_i(\phi) \to K_i(\neg K_i(\phi))$
\end{itemize}
The first three axioms present a minimal modal logic which captures valid formulas of all Kripke models. This axiomatisation is called modal 
system $\mathcal{K}$. The Truth axiom, also referred as $T$ axiom, states that whatever is known must be true. The last two axioms, also 
denoted by axiom $4$ and axiom $5$ respectively, express the introspective properties of an agent towards what it knows and what it doesn't 
know. The class of Kripke models with equivalence (accessibility) relations is denoted by $S$5. We would be working with beliefs, and not 
knowledge, \textit{where believed statements need not be true but must be internally consistent}. Constraining the Kripke structures with 
only serial, transitive, and Euclidean relations in $R$, allows us to talk about the beliefs of agents \cite{fagin2004reasoning}. In such a system, the axiom $T$ is replaced by axiom $D: B_i \phi \to \neg B_i \neg \phi$ (note the replacement of $K$ modality with $B$ modality) and 
the rest of the axioms remain the same with a belief modality $B$ replacing the knowledge modality $K$, and therefore, it is called as $KD45$ 
system. In the rest of the paper, we might use the terms knowledge and belief interchangeably but we confine ourselves only to belief as 
expressed in the $KD45$ system.

\section{Proposed Approach}
\begin{figure}[h]
\begin{center}
\includegraphics[height=1in, width=2.5in]{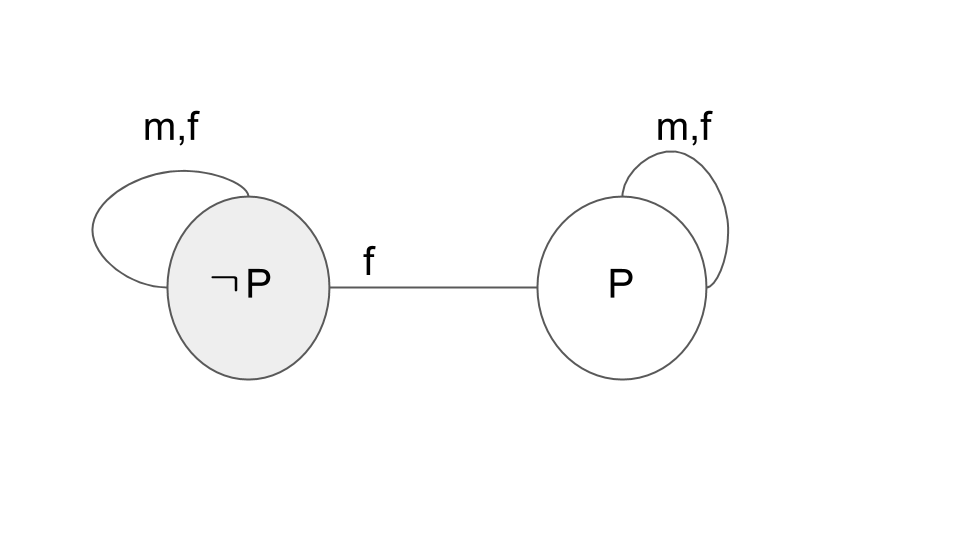}
\caption{An S5 Kripke model} 
\label{fig1}
\end{center}

\end{figure}

\textit{Informal Discussion:} When we model knowledge and beliefs in a multi-agent system using a Kripke model, the accessibility relations are defined for each agent for each state in the model to implicitly represent that all agents \textit{know} the other agents and it is common knowledge. The accessibility relations corresponding to an agent enables us to represent the respective agent's uncertainty about the real world but not about the agency. For example: Figure ~\ref{fig1} illustrates a \textit{S5} Kripke structure which captures the agents' beliefs (or uncertainty) about the truth of proposition \textit{p}.  
 
 It can be considered as a closed world representation (with respect to agents as well as propositions) given a finite set of propositions $P = \{p\}$ 
and a finite set of agents $\mathcal{AG} = \{m, f\}$. Let us assume that the shaded circle is the true world. From an external perspective we 
see that the model expresses that in the true state \textit{m} knows that p is false, \textit{f} is uncertain whether \textit{p} is true or 
false but \textit{f} is certain that \textit{m} knows \textit{whether} \textit{p}. Both the agents are \textit{implicitly aware} of each 
other. \newline We don't advocate an open world setting by making either \textit{P} or \textit{Ag} a non-finite set but we propose that 
agents should be allowed to expand their epistemic model, as and when updates about new agents joining in or existing ones leaving out, by 
applying the relations transformation functions we discuss in 
the following section. In this paper, we restrict ourselves to introducing new agents (and not new facts) in the system. The facts that agent 
can know (or believe) remain same. 

\subsection{Formalization}
Along the lines of the above definition of epistemic model, to represent an external as well as perspectival view of agents in a multi-agent system, we define some additional terms and notations as discussed below:\newline
 \textit{Local states corresponding to each agent -} $I:\mathcal{AG} \to 2^S$ is a function such that for all $i \in \mathcal{AG}$, the set $I(i) \subseteq S$ is the set of states that agent \textit{i} cannot distinguish from the real state in the initial model. It closely resembles the concept of \textit{designated states} discussed in \cite{bolander2011epistemic}. We assume that the system is initiated with a set of local states for each agent and we investigate how the set $I(i)$ corresponding to each agent \textit{i} evolves as new updates present themselves.\newline
 \textit{i-reachable states}: These are the states which can be reached by an \textit{i}-edge from any state.
We observe that $I(i)$ and its 1-hop neighborhood in the Kripke structure defines \textit{what i believes in} from a subjective perspective. 


Then there are \textit{j} edges emanating from states in this region that defines \textit{i}'s perception of \textit{j}'s beliefs.

To express an agent's uncertainty about the presence of another agent we consider the \textit{local states corresponding to the respective agent} and \textit{their neighborhood} and define two modal operators, $P_i$ read as \textit{Possibly\_an\_agent\_for\_i} and $C_i$ read as \textit{Certainly\_an\_agent\_for\_i}. 
\vskip 0.05in
 The full set of formulas can be given by the following BNF:
 \[\phi = p\, |\, \neg \phi\, |\, (\phi \wedge \varphi)\, |\, K_i\phi|\, C_i j|\, P_i j \] 
where $p \in P$ and $i,j \in \mathcal{AG}$

 The interpretation of these modalities is defined as follows:
\begin{itemize}

\item $(M,s) \models C_ij \text{ iff for all }t \text{ such that } (s,t) \in R(i),$ there exists atleast one $u$ such that $(t,u) \in R(j)$.

\item $M \models C_ij \text{ iff  for all } s \in I(i), (M,s) \models C_ij$.

\end{itemize}
We treat $P_i j$ as dual of $C_i j$ such that they mimic the `diamond' and the `box' operator respectively. 

We proceed with the \textit{possible worlds} semantics of epistemic logic for the language defined above. As in epistemic logic the possible worlds represent an agent's uncertainty in terms of possible epistemic alternatives, we extend it further to reason with an agent's uncertainty about the presence of other agents in terms the agents' accessibility to the shared epistemic alternatives.
\vskip 0.05in
 We expand on this concept with the help of the following example:

\begin{figure}[h]
\begin{center}
\includegraphics[height=1in, width=1.5in]{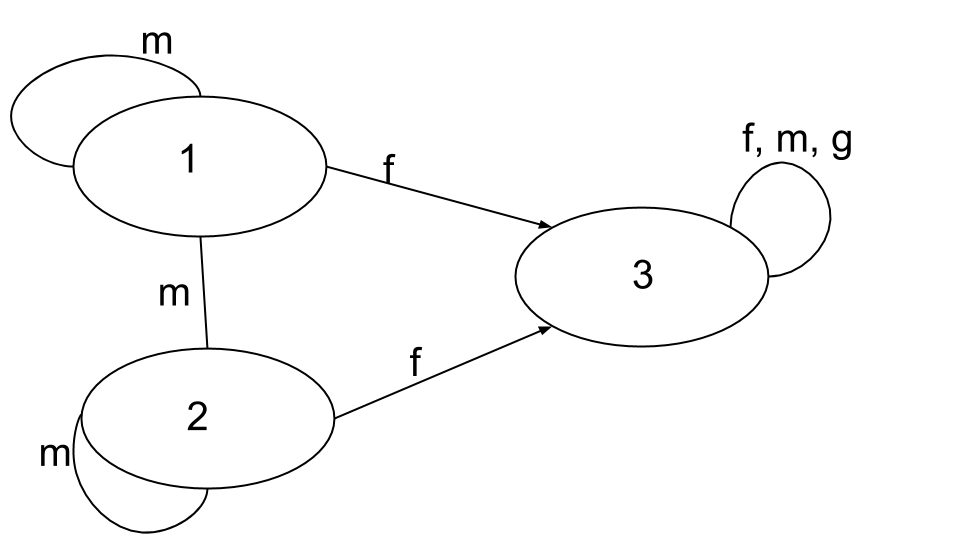}
\caption{A KD45 Kripke model}
\label{fig2}
\end{center}

\end{figure}
\vskip 0.05in

 Consider a KD45 Kripke model, in the figure below, defined over some finite set of propositions, a finite set of agents $\mathcal{AG} = \{m, 
f, g\}$  and a Kripke model $ M = \langle S,R,V \rangle $ where $S = \{1, 2, 3\}$, $R(m) = \{(1,1), (2,2), (2,1), (1,2), (3,3)\}$, $R(f) = \{(1,3), (2,3), (3,3)\}$, $R(g) = \{(3,3)\}$.
 Let the \textit{local states} for each agents be given as: $I(m) = \{1,2\}$, $I(f) = \{3\}$ and $I(g) = \{3\}$.\newline
\vskip 0.01in
 Let $(M, 1) \models p, (M, 2) \models \neg p$ and $(M, 3) \models p$. The epistemic formulas expressed on this model are: $\neg B_m p$, $B_m B_f p$, $B_f B_m p$, $B_f B_g p$ etc. 
%
The ontological formulas (as we have called them so) expressed on this model are: $\neg P_m g, B_m \neg P_m g$, $B_m B_f C_f g$, $B_f C_m g$ etc. 
\vskip 0.05in

The interpretation of the above model is that agent \textit{m}, which can not distinguish between the two epistemic states: \textit{1} and \textit{2} (represented by its local states), is certain of presence of only one agent \textit{f} and is itself oblivious of presence of another agent \textit{g}. In both of those epistemic states it believes that the agent \textit{f} considers a third and only epistemic state possible where it is not only certain of presence of \textit{m} but also certain of presence of another agent \textit{g}. Similarly, agent \textit{f} is certain of presence of agents \textit{m} as well as \textit{g} and believes that all three (including itself) are aware of each other. Clearly, if the true state had been $\{3\}$, we as an external agent can very well look at the model and tell that \textit{m} is mistaken in its beliefs about presence of only two agent in the system. Contrary to that, had one of $\{1, 2\}$ been the true state, we see that there are only two agents in the system which \textit{m} is aware of and it is also aware of the fact that \textit{f} is not only imagining an epistemic situation where another agent \textit{g} is present but also believes that everybody else shares the same belief.\newline
\vskip 0.05in
 One may argue that the \textit{seriality} of the relations, say $R(g)$ in the above example, vanishes. We emphasize again that 
the KD45 properties be maintained in a localised manner. Consider $R(m) = \{(1,1), (2,2), (1,2), (2,1), (3,3)\}$ which 
stands for the accessibility relation for \textit{m}. We don't use $R(m)$ to define \textit{m}'s subjective knowledge. Instead we use its 
subset $R'(m) = \{(1,1), (2,2), (1,2), (2,1)\}$ whose domain is restricted to $I(m)$. Similarly the subjective knowledge of \textit{f} and 
\textit{g} is defined using $R'(f) = \{(3,3)\}$ and $R'(g) = \{(3,3)\} $ respectively. 
Now that we can explicitly express the \textit{awareness} of agents about other agents, in the following section we introduce operators that 
can influence the same.

\subsection{Ontological updates}
In literature, update models from Dynamic Epistemic Logic \cite{van2007dynamic}  have been preferably used to formalize reasoning about information change in Kripke models. Another popular approach is using agent language \textit{mA+} \cite{baral2015action}. While we agree that the former can be used for modeling epistemic and ontic actions in our proposed setting too, it cannot be used to model the dynamics in agency that we discussed. The latter approach, agent language \textit{mA+} is known to be used with \textit{Finitary-S5 theories} \cite{son2014finitary} and therefore can not be used with arbitrary Kripke structures. 
To suit our purpose we describe two model transformation operators. We give their semantics using relation changing transition functions which transform one epistemic model to another. 
\begin{itemize}
\item \textit{Update\_offline(j)}: The updated model reflects that agent \textit{j} has (permanently) gone offline and other agents have updated their awareness about its absence and their beliefs about its beliefs. The resultant model, $M'$, can be constructed from the initial model $M$, in the following manner:
	\begin{itemize}
		\item Initialize the resultant model by creating a replica of the initial model: $M' = M$
		\item The set of agents with respect to the resultant model: $\mathcal{AG}'$ is set to $(\mathcal{AG} - j)$
		\item The local states $I(i)$ of all agents $i \in \mathcal{AG}'$ remains the same.
		\item Remove $R(j)$ such that for all $i \in \mathcal{AG}$, $\neg P_i j$ holds: $M' = M' \ominus^{r} R(j)$
	\end{itemize} 
\end{itemize} 	
	
Note that we use $\ominus^{r}$ operator to delete the specified relation from the model. Removal of edges may leave $M'$ disconnected. The regions that are not reachable from the local states of the agents are discarded.

\begin{figure}[h]
\begin{center}
\includegraphics[height=2in, width=3.5in]{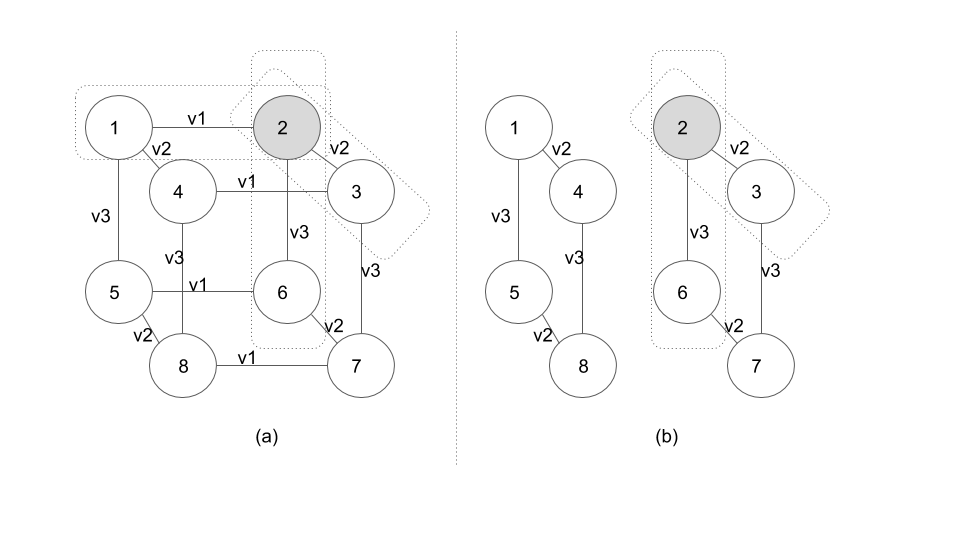}
\caption{An example scenario to demonstrate update\_offline(v1)}
\label{fig3} 
\end{center}

\end{figure}

 \textbf{\textit{Example:}} Consider a scenario in Figure ~\ref{fig3}a. There are three unmanned vehicles $\{v1, v2, v3\}$. The Kripke model illustrated in the figure\footnote{Note: This structure resembles the 3-muddy children puzzle but we do not model the same problem here. Consider it as an epistemic situation that shares similar neighborhood.} shows the eight possible worlds (each labeled node having some assignment over a finite set of propositions by the valuation function). The shaded node corresponds to the true world and the edges labeled with the agents represent the uncertainty of the respective agents. The dashed rectangles represent their local states. In this scenario, v1 leaves and the updated model is shown is Figure ~\ref{fig3}b. We observe that as v1 leaves, the epistemic possibilities that were present due to uncertainty of v1, gets disconnected and are of no relevance now. This component can be discarded.
 
\begin{itemize}
\item \textit{Update\_online(j, I(j))}: The updated model reflects that agent \textit{j} has joined the system. 
Other agents become aware of its presence. Besides that the earlier beliefs of the rest of the agents should remain intact. The model is updated with the local states of \textit{j}(as specified by the external modeler). The resultant model, $M'$, can be constructed from the initial model $M$, in the following manner:
	\begin{itemize}
		\item Initialize the resultant model by creating a replica of the initial model: $M' = M$
		\item The local states $I(i)$ of all agents $i \in \mathcal{AG}$ remains the same.
		\item The set of agents with respect to the resultant model: $\mathcal{AG}'$ is set to $(\mathcal{AG} + j)$
		\item Mark the local states (as specified in the update) for \textit{j}: $I(j)$ in $M'$.
		\item Add an equivalence relation $R(j)$ for \textit{j} on $S$, the set of all possible states: $ M' = M \oplus^{r} R(j)$ .
		 Now, for all $i \in\mathcal{AG}$, $C_i j$ holds true with respect to their respective $I(i)$s.
	\end{itemize} 	
Note that we use $\oplus^{r}$ operator to add the specified relation to the model. 
\end{itemize}

 As long as the agent updates are truthful and commonly known among all the agents, the model size (in terms of number of possible worlds) 
remain same and number of edges increases (or decreases) because we assume that the new agent cannot distinguish one epistemic state from others. The other agents too are aware of 
its ignorance. In planning based scenario, the beliefs of new agent (and hence that of the others too) can be further refined using information updates/requests in a goal driven manner. If we try to lift this assumption, say \textit{j} joins the group with some \textit{beliefs} then the 
updated model grows to accommodate the different beliefs of different agents about \textit{j}'s beliefs. For instance, each existing agent may 
falsely believe that the new agent shares the same view of the world that it itself has. If all the agents are 
biased to believe so, then the model expands atleast $|\mathcal{AG}|$ times.\newline


The updates that we discussed above bring common information for all the existing agents and the resultant model aligns with the true state of 
the world that the agents may still be uncertain about or yet to discover. But as it is the case with untruthful epistemic updates, that can be 
exploited to synthesize lies and deceptive plans, \textit{ontological updates} too can be 
exploited to synthesize \textit{ontological lies}. We give a brief account of ontological lies in the following section. 

\section{Epistemic lies versus Ontological lies}

If we take the view of subjective epistemic reasoning, we can define lying as follows. Lying is the communication of 
something that one does not believe in, usually done with the intention of misleading someone. We observe that there are two very different kinds of lies, requiring different kinds of cognitive machinery. The simpler kind of lie is 
\textit{epistemological}. Here the agent merely makes a statement that could have been true, but is not in fact. For example, a bridge player 
advertising a card she does not hold, or a person telling a habitual borrower that he has no money at hand. The second kind of lie is 
\textit{ontological} in which a new category is created; imagined and invented. For example, children are often told that a tooth fairy will come and take away a broken tooth. Though epistemological lies have been investigated by epistemic planning community, we, to the best of our beliefs, are not aware of such pursuits vis-a-vis ontological lies by artificial agents.

%
%
%

Our pursuit of studying ontological lies derives motivation from \textit{The Gruffalo} \cite{donaldson2016gruffalo}, written by Julia Donaldson it
is a children’s book featuring the
deceit carried out by a clever mouse, the leading character of the story, to safeguard itself from the
dangerous predators (a fox, an owl, a snake and finally, a gruffalo, \textit{a creature that the mouse thought
it was imagining}) in a forest. The interesting course of events that happens in the story is given
below:\footnote{$https://en.wikipedia.org/wiki/The\_Gruffalo$}\newline
\emph{The mouse, while taking a walk in a forest, runs into, one by one in sequence, a fox, an owl, and
a snake. Each of these animals, clearly intending to eat the mouse, invite him back to their home for
a meal. The cunning mouse declines each offer. To dissuade further advances, he tells each animal
that he has plans to dine with his friend, a gruffalo, a monster-like hybrid that is half grizzly bear
and half buffalo, whose favorite food happens to be the animal in question, and describes to each
the relevant dangerous features of the gruffalo's monstrous anatomy. Frightened that the gruffalo
might eat it, each animal flees. Knowing the gruffalo to be fictional, the mouse gloats thus:
\textit{Silly old fox/owl/snake, doesn't he know? there's no such thing as a gruffalo!}
After getting rid of the last animal, the mouse is shocked to encounter a real gruffalo - with all
the frightening features the mouse thought that he was inventing. The gruffalo threatens to eat the
mouse, but again the mouse resorts to imaginative deception. He tells the gruffalo that he, the
mouse, is the scariest animal in the forest. Laughing, the gruffalo agrees to follow the mouse as he
demonstrates how feared he is. The two walk through the forest, encountering in turn the animals
that had earlier menaced the mouse. Each is terrified by the sight of the pair and runs off - and
each time the gruffalo becomes more convinced about the fear that the mouse apparently evokes
in everyone. Observing the success of his deception, the mouse then threatens to make a meal
out of the gruffalo, which flees in haste, leaving the mouse to enjoy its vegetarian diet of a nut
in peace.}\newline
\vskip 0.05in
 As studied before \cite{singh2019planning} the first lie the mouse tells is of the latter type, while the second one is of the former type.
Ontological lies perhaps require more sophisticated cognitive machinery. Certainly a more
imaginative mind. We analyse \textit{ontological lies} using the setting discussed above. 


\subsection{Ontological lies: untruthful agent updates}
Let $(M, I(ag))$ and $(M', I'(ag))$ denote the initial and the resultant system respectively. Note that the agent update operators either add or remove accessibility relations and as a result, new epistemic states are also added to or removed from the initial model $M = \langle S, R, V \rangle$ get the resultant model $M' = \langle S', R', V' \rangle$. For ease of further discussion, we use $M(S)$ to denote the set of possible worlds in model $M$, $M(R)$ to denote all the relations in model, $M(V)$ to refer to the valuation function used and $I(M, ag)$ to refer to the set of local states of agent \textit{ag} in $M$. 

\subsubsection{Untruthful \textit{Update\_offline(j)} by agent i:} 
The updated model reflects that agent \textit{j} has (permanently) gone offline 
for the misinformed agents but not for \textit{i} and \textit{j}, that is, for all $k \in (\mathcal{AG} \setminus \{i,j\})$, $\neg P_k j$ holds where as for $k \in 
\{i,j\}$, $C_k j$ holds. The resultant model is constructed in the following manner:
	\begin{itemize}
		\item Given the initial model $M$, the resultant model $M'$ is initialized by creating two replicas of M, say $M^{act}$ and $M^{shift}$. $M^{act}$ corresponds to the true 
		scenario whereas $M^{shift}$ corresponds to epistemic state of the misinformed agents. We use $s^{act}$, $s^{shift}$ to 
		denote the states in $M^{act}$ and $M^{shift}$ corresponding to the state $s$ in the initial model $M$. Update the 
		domain of the resultant model $S' = M^{act}(S) \cup M^{shift}(S)$.
		
		 \item Remove the accessibility relations of the agent \textit{j}, which is falsely announced to be offline, from the region that reflects the 
 		epistemic region of misinformed agents: $M^{shift} = M^{shift} \ominus^{r} \, R^{shift}(j)$ where $R^{shift}(j) = \{(s
 		^{shift},\, t^{shift})$ $|(s,\, t) \in R(j)\}$. 
 		
 		 \item Remove accessibility of misinformed agents from the epistemic region that reflects the true state of affairs. $M^{act} = 
 		M^{act} \ominus^{r} \, R^{act}(k)$ where $R^{act}(k) = \{(s^{act},\, t^{act})|$ $(s,\, t) \in R(k)\}$ for all $k 
 		\in (\mathcal{AG}\setminus\{i,j\})$.
 		
 		 \item Update the accessibility relations of all the agents in the resultant model: firstly, for all $k \in \mathcal{AG}$, $R'(k) = 
 		 R_{act}(k) \oplus^{r} R_{shift}(k)$. Then add edges to the accessibility relations of misinformed agents such that their reachability goes from the 
 		 epistemic region that 
 		reflects the true state of affairs to the region that reflects the epistemic region of misinformed agents. It is done in the following manner: for all $k \in \mathcal{AG} 
 		\setminus\{i, j\}$, $R'(k) = R'(k) \, \oplus^{r} \, R_{addlinks}$ where $R_{addlinks} = $ $\{(s^{act}, t^{shift})|\, (s,\, t) \in R(k)\}$
 		
 		 \item Set the local states of the informed agents (\textit{i} and \textit{j}), as they were in the previous model $M$, in the 
 		 epistemic region represented by $M^{act}$. But the local states of misinformed agents (i.e. other than \textit{i} and \textit{j}) 
 		 can not lie in this region, they are shifted to the epistemic region represented by $M^{shift}$ which reflects the absence of 
 		 \textit{j}. Formally, in the resultant model, 
 		$M'$, for all $k \in (\mathcal{AG} \setminus \{i,j\})$, update $I'(k) = \{s^{shift}|\, s \in I(M, k)\}$ and for 
 		all $k \in \{i,j\}$, $I'(k) = \{s^{act}|\, s \in I(M, k)\}$.
		
		\item The above steps may leave the updated structure disconnected. Discard the components that are not reachable from the local 				states of the informed agents. The remaining structure is denoted as $M'$ and the set $S'$ and $R'$ are updated with $M'(S)$ and 				$M'(R)$ respectively.

\end{itemize}


\subsubsection{Untruthful \textit{Update\_online(j)} from by i:} 
We define similar relations transformation functions to update the model such that all the agents except \textit{i} believe that agent \textit{j} joins the group, that is, for all $k \in (\mathcal{AG} \setminus \{i\})$, $ C_k j$ holds while for \textit{i}, $\neg P_i j$ holds. The resultant model can be constructed in the following manner:

\begin{itemize}
		\item Given the initial model $M$, the resultant model $M'$ is initialized by creating two replicas of M, say $M^{act}$ and 					$M^{shift}$. $M^{act}$ corresponds to the true 
		scenario whereas $M^{shift}$ corresponds to epistemic state of the misinformed agents. We use $s^{act}$, $s^{shift}$ to 
		denote the states in $M^{act}$ and $M^{shift}$ corresponding to the state $s$ in the initial model $M$. Update the 
		domain of the resultant model $S' = M^{act}(S) \cup M^{shift}(S)$.
		
 		\item Add accessibility of the agent \textit{j}, which is falsely announced to have come online, to the region that reflects the 
 		epistemic region of misinformed agents by using {Update\_online(j, I(j))} on $M^{shift}$. 
 		
 		 \item Remove accessibility of misinformed agents from the epistemic region that reflects the true state of affairs. $M^{act} = 
 		M^{act} \ominus^{r} \, R^{act}(k)$ where $R^{act}(k) = \{(s^{act},\, t^{act})|$ $(s,\, t) \in R(k)\}$ for all $k 
 		\in (\mathcal{AG}\setminus\{i\})$.
 		
 		 \item Update the accessibility relations of all the agents in the resultant model: firstly, for all $k \in \mathcal{AG}$, $R'(k) = R_{act}
 		 (k) \oplus^{r} R_{shift}(k)$. Then, add edges to the accessibility relations of misinformed agents such that their reachability goes from the 
 		 epistemic region that 
 		reflects the true state of affairs to the region that reflects the epistemic region of misinformed agents: for all $k \in 
 		(\mathcal{AG} 
 		\setminus\{i\})$, $R'(k) = R'(k) \, \oplus^{r} \, R_{addlinks}$ where $R_{addlinks} = $ $\{(s^{act}, t^{shift})|\, (s,\, t) \in R(k)\}$.
 		

		 \item Set the local states of the informed agent (\textit{i}), as they were in the previous model $M$, in the 
 		 epistemic region represented by $M^{act}$. But the local states of misinformed agents (i.e. other than \textit{i}) 
 		 can not lie in this region, they are shifted to the epistemic region represented by $M^{shift}$ which reflects the presence of 
 		 \textit{j}. Formally, in the resultant model, 
 		$M'$, for all $k \in (\mathcal{AG} \setminus \{i\})$, update $I'(k) = \{s^{shift}|\, s \in I(M, k)\}$ and for 
 		all $i$, $I'(i) = \{s^{act}|\, s \in I(M, i)\}$.
		
		\item The above steps may leave the updated structure disconnected. Discard the components that are not reachable from the local 				states of the informed agents. The remaining structure is denoted as $M'$ and the set $S'$ and $R'$ are updated with $M'(S)$ and 				$M'(R)$ respectively.

\end{itemize}

\begin{figure}[h]
\begin{center}

\includegraphics[height=2in, width=3.5in]{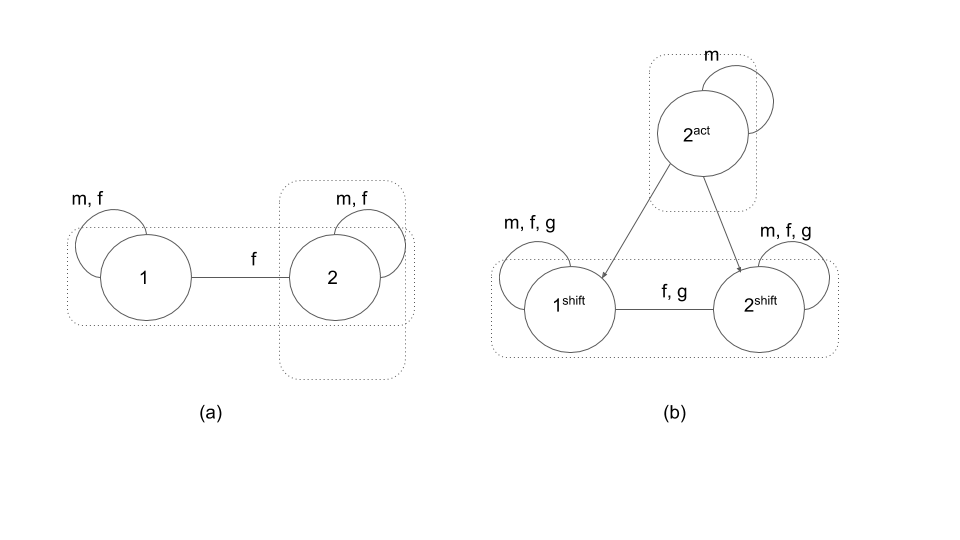}
\caption{An example scenario to demonstrate untruthful update\_online(g)}
\label{fig4} 

\end{center}
\end{figure}

\textbf{\textit{Example:}} In Figure ~\ref{fig4} we demonstrate the model transformation from ~\ref{fig4}a to ~\ref{fig4}b owing to the false ontological update by the mouse (m) about an imaginary gruffalo (g) to the fox (f) as discussed in the story earlier. The dashed rectangular boxes show their initial and shifted local states in ~\ref{fig4}a and ~\ref{fig4}b respectively.
 
\section{Conclusion}
In this paper we explore an epistemic modeling technique based on Kripke structures wherein agents may be able to influence other agents' mental states by (mis)informing them about the presence/absence of other agents. We define two modal operators to express an agent's certainty about the agency. Then we define model transformation operators which we call ontological updates. We observe that the model grows faster in case of untruthful updates. We also discuss some examples to demonstrate the working of our approach. We feel that these ontological updates can be used in planning based settings in artificial intelligence. The idea of \textit{imagining fictional characters} seems to be an in interesting pursuit if studied in an epistemic planning based setting augmented with ontological updates. We hope to explore this avenue in the future. In this formalism, we restrict ourselves to the propositions that stand for world knowledge thereby excluding the agent-specific knowledge. For instance, in the muddy children puzzle each proposition stands for some child being muddy. It is not clear how to maintain and handle these propositions and hence beliefs about them in the setting where the agents may join or leave the system. These are the few issues that the discussed approach does not address but need to be addressed in order for the approach to be useful for solving a wide class of problems.

\bibliographystyle{plain}  
\bibliography{template}  


\end{document}